\documentclass{lmcs} 
\pdfoutput=1

\usepackage[utf8]{inputenc}

\usepackage{lastpage}
\lmcsdoi{17}{1}{10}
\lmcsheading{}{\pageref{LastPage}}{}{}%
{Dec.~16,~2019}{Feb.~03,~2021}{}


\usepackage{xspace}

\newcommand{\G}{{\mathcal G}}

\renewcommand{\P}{{\mathcal P}}
\newcommand{\M}{{\mathcal M}}

\newcommand{\PO}{Player~$1$\xspace}
\newcommand{\PT}{Player~$2$\xspace}
\newcommand{\PLi}{Player~$i$\xspace}
\renewcommand{\ni}{- \! i\xspace}
\newcommand{\PLni}{Player~$\ni$\xspace}
\newcommand{\thresh}{\texttt{Thresh}\xspace}
\newcommand{\stam}[1]{}
\newcommand{\zug}[1]{\langle #1  \rangle}
\newcommand{\set}[1]{\{ #1 \}}

\newcommand{\play}{\text{play}\xspace}
\newcommand{\Buchi}{\text{B\"uchi}\xspace}

\newcommand{\Muller}{\text{M\"uller}\xspace}

\newcommand{\Nat}{\mathbb{N}}

\begin{document}

\title{Determinacy in Discrete-Bidding Infinite-Duration Games}
\titlecomment{{\lsuper*} A preliminary version of this paper appeared in the proceedings of the 30th CONCUR, LIPIcs 140, pages 20:1--20:17, Schloss Dagstuhl, 2019. This research was supported in part by the Austrian Science Fund (FWF) under grants S11402-N23 (RiSE/SHiNE), Z211-N23 (Wittgenstein Award), and M 2369-N33 (Meitner fellowship).}

\author[M.~Aghajohari]{Milad Aghajohari\rsuper{a}}	
\address{\lsuper{a}Sharif University of Technology, Iran}	
\email{mi.aghajohari@gmail.com}  

\author[G.~Avni]{Guy Avni\rsuper{b}}	

\author[T.A.~Henzinger]{Thomas A. Henzinger\rsuper{b}}	
\address{\lsuper{b}IST Austria}	
\email{gguyavni@gmail.com}  
\email{tah@ist.ac.at}  





\begin{abstract}
 In two-player games on graphs, the players move a token through a graph to produce an infinite path, which determines the winner of the game. Such games are central in formal methods since they model the interaction between a non-terminating system and its environment. In bidding games the players bid for the right to move the token: in each round, the players simultaneously submit bids, and the higher bidder moves the token and pays the other player. Bidding games are known to have a clean and elegant mathematical structure that relies on the ability of the players to submit arbitrarily small bids. Many applications, however, require a fixed granularity for the bids, which can represent, for example, the monetary value expressed in cents. We study, for the first time, the combination of {\em discrete-bidding} and {\em infinite-duration} games. Our most important result proves that these games form a large determined subclass of concurrent games, where {\em determinacy} is the strong property that there always exists exactly one player  who can guarantee winning the game. In particular, we show that, in contrast to non-discrete bidding games, the mechanism with which tied bids are resolved plays an important role in discrete-bidding games. We study several natural tie-breaking mechanisms and show that, while some do not admit determinacy, most natural mechanisms imply determinacy for every pair of initial budgets.
 \end{abstract}

\maketitle

\section{Introduction}
Two-player infinite-duration games on graphs are a central class of games in formal verification~\cite{AG11} and have deep connections to foundations of logic \cite{Rab69}. They are used to model the interaction between a system and its environment, and the problem of synthesizing a correct system then reduces to finding a winning strategy in a graph game \cite{PR89}. A graph game proceeds by placing a token on a vertex in the graph, which the players move throughout the graph to produce an infinite path (``play'') $\pi$. The winner of the game is determined according to $\pi$. 

Two ways to classify graph games are according to the type of {\em objectives} of the players, and according to the {\em mode of moving} the token. For example, in {\em reachability games}, the objective of \PO is to reach a designated vertex $t$, and the objective of \PT is to avoid $t$. An infinite play $\pi$ is winning for \PO iff it visits $t$. The simplest mode of moving is {\em turn based}: the vertices are partitioned between the two players and whenever the token reaches a vertex that is controlled by a player, he decides how to move the token.

In {\em bidding} games, in each turn, a bidding takes place to determine which player moves the token. Bidding games were introduced in \cite{LLPSU99,LLPU96}, where the main focus was on a concrete bidding rule, called {\em Richman} rule (named after David Richman), which is as follows: Each player has a budget, and before each move, the players simultaneously submit bids, where a bid is legal if it does not exceed the available budget. The player who bids higher wins the bidding, pays the bid to other player, and moves the token. 

Bidding games exhibit a clean and elegant theory. The central problem that was previously studied concerned the existence of a necessary and sufficient {\em threshold budget}, which allows a player to achieve his objective. Assuming the sum of budgets is $1$, the threshold budget at a vertex $v$, denoted $\thresh(v)$, is such that if \PO's budget exceeds $\thresh(v)$, he can win the game, and if \PT's budget exceeds $1-\thresh(v)$, he can win the game. Threshold budgets are known to exist in bidding reachability games \cite{LLPSU99,LLPU96} with variants of the first-price bidding rule that is described above. Only reachability Richman-bidding games, however, are equivalent to {\em random-turn games} \cite{PSSW09}, which are a special case of {\em stochastic games} \cite{Con90} in which in each turn, the player who moves is chosen according to a probability distribution. Interestingly, a more general and robust equivalence with random-turn games holds for mean-payoff bidding games, which are infinite-duration games, with Richman bidding \cite{AHC19}, {\em poorman} bidding \cite{AHI18}, which are similar to Richman bidding except that the winner of a bidding pays the ``bank'' rather than the other player, and {\em taxman} bidding \cite{AHZ19}, which span the spectrum between Richman and poorman bidding. 

These theoretical properties of bidding games highly depend on the ability of the players to submit arbitrarily small bids. Indeed, in poorman games, the bids tend to $0$ as the game proceeds. Even in Richman reachability games, when the budget of \PO at $v$ is $\thresh(v) + \epsilon$, a winning strategy bids so that the budget always exceeds the threshold budget and, either the game is won or \PO's surplus, namely the difference between his budget and the threshold budget, strictly increases. This strategy uses bids that are exponentially smaller than $\epsilon$.

For practical applications, however, allowing arbitrary granularity of bids is unreasonable. For example, in formal methods, graph games are used to reason about multi-process systems, and bidding naturally models ``scrip'' systems, which use internal currency in order to prioritize processes. Car-control systems are one example, where different components might send conflicting actions to the engine, e.g., the cruise control component can send the action ``accelerate'' while the traffic-light recognizer can send ``stop''. Bidding then specifies the level of criticality of the actions, yet for this mechanism to be practical, the number of levels of criticality (bids) must stay small. Bidding games can be used in settings in which bids represent the monetary value of choosing an action. Such settings typically have a finite granularity, e.g., cents. One such setting is Blockchain technology \cite{CGV18,ABC16}, where players represent agents that are using the service, and their bids represent transaction fees to the miners. A second such setting is reasoning about ongoing auctions like the ones used in the internet for advertisement allocation \cite{Mut09}. Bidding games can be used to devise bidding strategies in such auctions. Motivation for bidding games also comes from recreational games, e.g., bidding chess \cite{BP09} or tic-tac-toe\footnote{\url{http://biddingttt.herokuapp.com/}}, where it is unreasonable for a human player to keep track of arbitrarily small and possibly irrational numbers.

In this work, we study {\em discrete-bidding games} in which the granularity of the bids is restricted to be natural numbers. A key difference from the continuous-bidding model is that there, the issue of tie breaking was largely ignored, which is possible since one can consider cases where the initial budget does not equal $\thresh(v)$. In discrete-bidding, however, ties are a central part of the game. A discrete-bidding game is characterized explicitly by a tie-breaking mechanism in addition to the standard components, i.e.,  an arena, the players' budgets, and an objective. We investigate several tie-breaking mechanisms and show how they affect the properties of the game. Discrete-bidding games with reachability objectives were first studied in \cite{DP10}. The focus in that paper was on extending the Richman theory to the discrete domain, and we elaborate on their results later in this section.

A central concept in game theory is a {\em winning strategy}: a strategy that a player can reveal before the other player, and still win the game. A game is {\em determined} if exactly one of the players can guarantee winning the game. The simplest example of a non-determined game is a two-player game called {\em matching pennies}: Each player chooses $1$ (``heads'') or $0$ (``tails''), and \PO wins iff the parity of the sum of the players' choices is $0$. Matching pennies is not determined since if \PO reveals his choice first, \PT will choose opposite and win the game, and dually for \PT. 

Discrete-bidding games are a subclass of {\em concurrent} graph games \cite{AHK02}, in which in each turn, the players simultaneously select actions, and the joint vector of actions determines the next position. A bidding game $\G$ is equivalent to a concurrent game $\G'$ that is played on the ``configuration graph'' of $\G$: each vertex of $\G'$ is a tuple $\zug{v, B_1, B_2, s}$, where $v$ is the vertex in $\G$ on which the token is situated, the players' budgets are $B_1$ and $B_2$, and $s$ is the state of the tie-breaking mechanism. An action in $\G'$ corresponds to a bid and a vertex to move to upon winning the bidding. Concurrent games are not in general determined since matching pennies can be modelled as a concurrent game.

\begin{figure}[t]
\centering
\includegraphics[height=3cm]{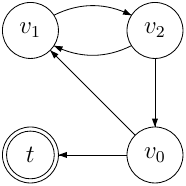}
\caption{A bidding game that is not determined with alternating tie-breaking, when the initial configuration is $\zug{v_0, 1, 1^*}$.}
\label{fig:alternate}
\end{figure}

The central question we address in this work asks under which conditions bidding games are determined. We show that determinacy in bidding games highly depends on the tie-breaking mechanism under use. We study natural tie-breaking mechanisms, show that some admit determinacy while others do not. The simplest tie-breaking rule we consider alternates between the players: \PO starts with the {\em advantage}, when a tie occurs, the player with the advantage wins, and the advantage switches to the other player. We show that discrete-bidding games with alternating tie-breaking are not determined, as we demonstrate below. 
\begin{exa}
\label{ex:alternate}
Consider the bidding reachability game that is depicted in Fig.~\ref{fig:alternate}. We depict the player who has the advantage with a star. We claim that no player has a winning strategy when the game starts from the configuration $\zug{v_0, 1, 1^*}$, thus the token is placed on $v_0$, both budgets equal $1$, and \PT has the tie-breaking advantage. We start by showing that if \PT reveals his first bid before \PO, then \PO can guarantee winning the game. There are two cases. First, if \PT bids $0$, \PO bids $1$ and draws the game to~$t$. Second, if \PT bids $1$, then \PO bids $0$, and the game reaches the configuration $\zug{v_1, 2, 0^*}$. Next, both players bid $0$ and we reach $\zug{v_2, 2^*, 0}$. \PO wins by bidding $1$ twice; indeed, the next two configurations are $\zug{v_0, 1^*, 1}$ and either $\zug{t, 0, 2^*}$, if \PT bids $1$, or $\zug{t, 0^*, 2}$, if he bids $0$. The proof that \PO loses when he reveals his first bid before \PT can be found in Theorem~\ref{thm:alternate}.\hfill\qed
\end{exa}

We generalize the alternating tie-breaking mechanism as follows. A {\em transducer} is similar to an automaton only that the states are labeled by output letters. In {\em transducer-based} tie breaking, a transducer is run in parallel to the game. The transducer reads information regarding the biddings and outputs which player wins in case of a tie. Alternating tie-breaking is a special case of transducer tie-breaking in which the transducer is a two-state transducer, where the alphabet consists of the letters $\top$ (``tie'') and $\bot$ (``no-tie'') and the transducer changes its state only when the first letter is read. 

\begin{figure}[ht]
\center
\includegraphics[height=1.5cm]{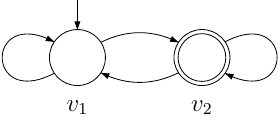}
\hspace{1cm}
\includegraphics[height=1.5cm]{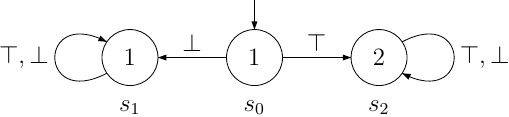}
\caption{On  the left, a \Buchi game that is not determined when tie-breaking is determined according to the transducer on the right, where the letters $\top$ and $\bot$ respectively represent ``tie'' and ``no tie''.}
\label{fig:buchi-not-det}
\end{figure}

\begin{exa}
\label{ex:Buchi-non-det}
We describe another simpler game that is not determined. In a {\em \Buchi game}, \PO wins a play iff it visits an accepting state infinitely often. Consider the \Buchi bidding game that is depicted on the left of Fig.~\ref{fig:buchi-not-det} with the tie-breaking uses the transducer on the right of the figure. That is, if a tie occurs in the first bidding, \PT wins all ties for the rest of the game, and otherwise \PO wins all ties. Note that for $i \in \set{1,2}$, no matter what the budgets are, if \PLi wins all ties, he can win the game. A winning strategy for \PLi always bids $0$. Intuitively, the other player must invest a unit of budget for winning a bidding and leaving $v_i$, thus the game eventually stays in $v_i$. So, the winner of the game is determined according to the outcome of the first bidding. Suppose both players' initial budgets are positive and \PT's budget is not larger than \PO's, thus \PT cannot force a win in the first bidding. Then, the players essentially play a matching-pennies game in the first round, hence no player has a winning strategy.\hfill\qed
\end{exa}

We proceed to describe our positive results. For transducer-based tie-breaking, we show that bidding games are determined when the transducer is un-aware of the occurrence of ties. Note that this property of the transducer is also a necessary to ensure determinacy since the transducer in Example~\ref{ex:Buchi-non-det} is aware of ties. The second tie-breaking mechanism for which we show determinacy is {\em random tie-breaking}: a tie is resolved by tossing a coin that determines the winner of the bidding. Finally, a tie-breaking mechanism that was introduced in \cite{DP10} is advantage based, except that when a tie occurs, the player with the advantage can choose between (1) winning the bidding and passing the advantage to the other player, or (2) allowing the other player to win the bidding and keeping the advantage. Determinacy for reachability games with this tie-breaking mechanism was shown in \cite{DP10}. The technique that is used there cannot be extended to the other tie-breaking mechanisms we study. We show an alternative proof for advantage-based tie-breaking and extend the determinacy result for richer objectives beyond reachability. 

We obtain our positive results by developing a unified proof technique to reason about bidding games, which we call {\em local determinacy}. Intuitively, a concurrent game is locally determined if from each vertex, there is a player who can reveal his action before the other player. We show that locally-determined reachability games are determined and then extend to {\em \Muller} games, which are richer qualitative games. We expect our technique to extend to show determinacy in other fragments of concurrent games unlike the technique in \cite{DP10}, which is tailored for bidding games. 

Determinacy has computational complexity implications; namely, finding the winner in a bidding game  with objective $\alpha$ when the budgets are given in unary is as hard as solving a turn-based game with objective $\alpha$, and we show a simple reduction in the other way for bidding games. Finally, we establish results for strongly-connected discrete-bidding games.

\section{Preliminaries}
\subsection{Concurrent and turn-based games}
A {\em concurrent} game is a two-player game that is played by placing a token on a graph. In each turn, both players simultaneously select actions, and the next vertex the token moves to is determined according to their joint actions. The players' actions give rise to an infinite path $\pi$ in the graph. A game is accompanied by an objective for \PO, who wins iff $\pi$ meets his objective. We specify standard objectives in games later in the section. For $i \in \set{1,2}$, we use $\ni$ to refer to the other player, namely $\ni = 3-i$.

Formally, a concurrent game is played on an arena $\zug{A, V, \lambda, \delta}$, where $A$ is a finite non-empty set of actions, $V$ is a finite non-empty set of vertices, the function $\lambda: V \times \set{1,2} \rightarrow 2^A \setminus \emptyset$ specifies the allowed actions for \PLi in vertex $v$, and $\delta: V \times A \times A \rightarrow V$ specifies, given the current vertex and a choice of actions for the two players, the next vertex the token moves to. We call $u \in V$ a {\em neighbor} of $v \in V$ if there is a pair of allowed action $a^1,a^2 \in A$ at $v$ with $u = \delta(v, a^1, a^2)$. We use $N(v) \subseteq V$ to denote the set of neighbors of $v$. We say that \PLi \ {\em controls} a vertex $v \in V$ if his actions uniquely determine where the token proceeds to from $v$. That is, for every $a \in \lambda(v,i)$ there is a vertex $u$ such that, for every allowed action $a'$ of \PLni, we have $\delta(v, a, a') = u$. A {\em turn-based} game is a special case of a concurrent game in which each vertex is controlled by one of the players.

\subsection{Bidding games}
A (discrete) {\em bidding game} is a special case of a concurrent game. The game is played on a graph and both players have budgets. In each turn, a bidding takes place to determine which player gets to move the token. Formally, a bidding game is played on an arena $\zug{V, E, N, \M}$, where $V$ is a set of vertices, $E \subseteq (V \times V)$ is a set of edges, $N \in \Nat$ represents the total budget, and the {\em tie-breaking mechanism} is $\M$ on which we elaborate below. 

We formalize the semantics of a bidding game $\G = \zug{V, E, N, \M}$ by means of a concurrent game $\zug{A, V', \lambda, \delta}$. For ease of presentation, in a vertex that is controlled by one player, we list only the neighboring vertices rather than specifying the allowed actions. The set of actions correspond to the possible bids, thus $A = \set{0,\ldots,N}$. The vertices are partitioned between {\em configuration} vertices and {\em intermediate} vertices. Intuitively, biddings occur in configuration vertices. Intermediate vertices are convenient for ``book keeping''; the winner chooses the successor vertex and the state of the tie-breaking mechanism is updated. Formally, a configuration vertex is $c=\zug{v, B_1, B_2 , s}$, where $v \in V$ is the vertex on which the token is placed on in the bidding game $\G$, for $i \in \set{1,2}$, the budget of \PLi is $B_i \in \set{0,\dots, N}$, where $B_1 + B_2 = N$, and $s$ is the state of the tie-breaking mechanism as we elaborate below. The set of allowed actions in $c$ is $\set{0,\ldots, B_i}$ for \PLi, which, again, corresponds to the legal bids. 

An intermediate vertex is $x = \zug{c, b_1, b_2}$, where $c = \zug{v, B_1, B_2, s}$ is a configuration vertex and $b_i \in \set{0,\ldots, N}$, for $i \in \set{1,2}$. The neighbors of a configuration vertex $c$ are of the form $\zug{c, b_1, b_2}$, for every pair of allowed actions $b_1$ and $b_2$ for the two players in $c$. Let $b_1, b_2 \in \set{0,\ldots, N}$. Suppose $b_1 > b_2$ and the case of $b_2 > b_1$ is dual. \PO wins the bidding at $c$. Let $B'_1 = B_1 - b_1$ and $B'_2 = B_2 + b_2$, thus \PO pays \PT the winning bid. \PO controls the intermediate vertex $x$. Its neighbors are of the form $\zug{v', B'_1, B'_2, s'}$, where $v'$ is a neighbor of $v$ in $\G$ and $s'$ is the updated tie-breaking mechanism as we elaborate below.

We proceed to the case of ties, i.e., when $b_1 = b_2$, and describe three types of tie-breaking mechanisms. 
\begin{description}
  \item[Transducer-based] A {\em transducer} is $T = \zug{\Sigma, Q, q_0, \Delta, \Gamma}$, where $\Sigma$ is a set of letters, $Q$ is a set of states, $q_0 \in Q$ is an initial state, $\Delta: Q \times \Sigma \rightarrow Q$ is a partial deterministic function, and $\Gamma: Q \rightarrow \set{1,2}$ is a labeling of the states. Intuitively, $T$ is run in parallel to the bidding game and its state is updated according to the outcomes of the biddings. Whenever a tie occurs and $T$ is in state $s \in Q$, the winner of the bidding is $\Gamma(s)$. The information according to which tie-breaking is determined is represented by the alphabet of $T$. In general, the information can include the vertex on which the token is located and the result of the previous bidding, i.e., the winner, whether or not a tie occurred, and the winning bid, thus $\Sigma = V \times \set{1,2} \times \set{\bot, \top} \times \Nat$. 
  \item [Random-based] A tie is resolved by choosing the winner uniformly at random.
  \item [Advantage-based] Exactly one player holds the {\em advantage}. Suppose \PLi holds the advantage and a tie occurs. Then \PLi chooses who wins the bidding. If he calls the other player the winner, \PLi keeps the advantage, and if he calls himself the winner, the advantage switches to the other player. 
\end{description}
We describe the updates to the tie-breaking mechanism's state when using the three mechanisms above. Consider a configuration $c = \zug{v, B_1, B_2 , s}$ and an intermediate vertex $\zug{c, b_1, b_2}$. With transducer-based mechanism, the state $s$ is a state in the transducer $T$. If $b_1 \neq b_2$, the player who controls $\zug{v, b_1, b_2}$ is determined as in the above. In case $b_1 = b_2$, then Player~$\Gamma(s)$ controls the vertex. In both cases, we update the state of the tie-breaking mechanism by feeding it the information on the last bidding; who won, whether a tie occurred, and what vertex the winner chose, thus we set $s' = \Delta(s, \sigma)$, where $\sigma = \zug{v', i, \bot, b_i}$ in case \PLi wins the bidding with his bid of $b_i$, moves to $v'$, and no tie occurs. The other cases are similar. 

In random-based tie-breaking, the mechanism has no state, thus we can completely omit $s$. Consider an intermediate vertex $\zug{c, b_1,b_2}$. The case of $b_1\neq b_2$ is as in the above. Suppose both players bid $b$. For ease of presentation we assume $b >0$, and the case of $b=0$ is defined in a similar manner. The intermediate vertex $\zug{c, b, b}$ is controlled by ``Nature''. It has two probabilistic outgoing transitions; one transition leads to the intermediate vertex $\zug{c, b, b-1}$, which represents \PO winning the bidding with a bid of $b$, and the other to the intermediate vertex $\zug{c, b-1, b}$, which represents \PT winning the bidding with a bid of $b$. We elaborate on the semantics of concurrent games with probabilistic edges in Section~\ref{sec:stochas}.

Finally, in advantage-based tie-breaking, the state of the mechanism represents which player has the advantage, thus $s \in \set{1,2}$. Consider an intermediate vertex $\zug{c, b_1,b_2}$. When a tie does not occur, there is no need to update $s$. When $b_1 = b_2$,  then Player~$s$ controls $\zug{c,b_1, b_2}$ and the possibility to choose who wins the bidding. Choosing to lose the bidding is modelled by no update to $s$ and moving to an intermediate vertex that is controlled by Player~$-s$ from which he chooses a successor vertex and the budgets are updated accordingly. When Player~$s$ chooses to win the bidding we proceed directly to the next configuration vertex, update the budgets, and the mechanism's state to $3-s$.

\subsection{Strategies, plays, and objectives}
A {\em strategy} is, intuitively, a recipe that dictates the actions that a player chooses in a game. Formally, a finite {\em history} of a concurrent game is a sequence  $\zug{v_0,a^1_0, a^2_0},\ldots, \zug{v_{n-1},a^1_{n-1},a^2_{n-1}}, v_n \in (V \times A \times A)^*\cdot V$ such that, for each $0 \leq i <n$, we have $v_{i+1} = \delta(v_i, a^1_i, a^2_i)$. A strategy is a function from $(V \times A \times A)^*\cdot V$ to $A$. We restrict attention to legal strategies that assign only allowed actions, thus for every history $\pi \in (V \times A \times A)^*\cdot V$ that ends in $v \in V$, a legal strategy $\sigma_i$ for \PLi has $\sigma_i(\pi) \in \lambda(v, i)$. Two strategies $\sigma_1$ and $\sigma_2$ for the two players and an initial vertex $v_0$, determine a unique {\em play}, denoted $\play(v_0, \sigma_1, \sigma_2) \in (V \times A \times A)^\omega$, which is defined as follows. The first element of $\play(v_0, \sigma_1, \sigma_2)$ is $\zug{v_0, \sigma_1(v_0), \sigma_2(v_0)}$. For $i \geq 1$, let $\pi^i$ denote the prefix of length $i$ of $\play(v_0, \sigma_1, \sigma_2)$ and suppose its last element is $\zug{v_i, a^1_i, a^2_i}$. We define $v_{i+1} = \delta(v_i, a^{i}_1, a^{i}_2)$,  $a^{i+1}_1 = \sigma_1(\pi^i \cdot v_{i+1})$, and $a^{i+1}_2 = \sigma_2(\pi^i\cdot v_{i+1})$. The {\em path} that corresponds to $\play(v_0, \sigma_1, \sigma_2)$ is $v_0,v_1,\ldots$. 

An {\em objective} for \PO is a subset of infinite paths $\alpha \subseteq V^\omega$. We say that \PO wins $\play(v_0, \sigma_1, \sigma_2)$ iff the path $\pi$ that corresponds to $\play(v_0, \sigma_1, \sigma_2)$ satisfies the objective, i.e., $\pi \in \alpha$. Let $inf(\pi) \subseteq V$ be the subset of vertices that $\pi$ visits infinitely often. We consider the following objectives. 
\begin{description}
  \item [Reachability] A game is equipped with a target set $T \subseteq V$. A play $\pi$ is winning for \PO, the reachability player, iff it visits $T$.
  \item [\Buchi{}] A game is equipped with a set $T \subseteq V$ of accepting vertices. A play $\pi$ is winning for \PO iff it visits $T$ infinitely often.
  \item [Parity]  A game is equipped with a function $p: V \rightarrow \set{1,\ldots, d}$, for $d \in \Nat$. A play $\pi$ is winning for \PO iff $\max_{v \in inf(\pi)} p(v)$ is odd.
  \item [\Muller{}] A game is equipped with a set $T \subseteq 2^V$. A play $\pi$ is winning for \PO iff $inf(\pi) \in T$. 
\end{description}

\section{A Framework for Proving Determinacy}
\subsection{Determinacy}
{\em Determinacy} is a strong property of games, which intuitively says that exactly one player has a winning strategy. That is, the winner can reveal his strategy before the other player, and the loser, knowing how the winner plays, still loses. 

\begin{defi}[Determinacy]
A strategy $\sigma_i$ is a {\em winning strategy} for \PLi at vertex $v$ iff for every strategy $\sigma_{\ni}$ for \PLni, \PLi wins $\play(v, \sigma_1, \sigma_2)$. We say that a game $\zug{V, E, \alpha}$ is {\em determined} if from every vertex $v \in V$ either \PO has a winning strategy from $v$ or \PT has a winning strategy from $v$. 
\end{defi}

While concurrent games are not determined (e.g., ``matching pennies''), turn-based games are largely determined.
\begin{thmC}[\cite{Mar75}]
Turn-based games with objectives that are Borel sets are determined. In particular, turn-based \Muller games are determined.
\end{thmC}

We describe an alternative definition for determinacy in concurrent games. Consider a concurrent game $\G = \zug{A, V, \lambda, \delta, \alpha}$. Recall that in $\G$, in each turn, the players simultaneously select an action, and their joint actions  determine where the token moves to. For $i \in \set{1,2}$, let $\G_i$ be the turn-based game that, assuming the token is placed on a vertex $v$, \PLi selects an action first, then \PLni selects an action, and the token proceeds from $v$ as in $\G$ given the two actions. Formally, the game $\G_1$ is a turn-based game $\zug{A, V \cup (V \times A), \lambda', \delta', \alpha'}$, and the definition for $\G_2$ is dual. The vertices that are controlled by \PO are $V_1 = V$ and $V_2 = V \times A$. For $v \in V$, we have $\lambda'(v, 1) = \lambda(v, 1)$ and since \PO controls $v$, we arbitrarily fix $\lambda'(v, 2) = A$. For $a_1 \in \lambda(v, 1)$ and $a_2 \in A$, we define $\delta(v, a_1, a_2) = \zug{v, a_1}$. Similarly, we define $\lambda'(\zug{v, a_1}, 1) = A$ and $\lambda'(\zug{v, a_1}, 2) = \lambda(v, 2)$. For $a_1' \in A$ and $a_2 \in \lambda(v, 2)$, we define $\delta'(\zug{v, a_1}, a_1', a_2) = \delta(v, a_1, a_2)$.  Finally, an infinite play $v_1, \zug{v_1, a_1}, v_2, \zug{v_2, a_2},\ldots,$ is in $\alpha'$ iff $v_1, v_2,\ldots$ is in $\alpha$. Recall that in bidding games, intermediate vertices are controlled by one player and the only concurrent moves occur when revealing bids. Thus, when $\G$ is a bidding game, in $\G_i$, \PLi always reveals his bids before \PLni.

\begin{prop}
\label{prop:det}
A strategy $\sigma_i$ is winning for \PLi in $\G$ at vertex $v$ iff it is winning in $\G_i$ from $v$. Then, $\G$ is determined at $v$ iff either \PO wins in $\G_1$ from $v$ or \PT wins in $\G_2$ from $v$.
\end{prop}

\subsection{Local and global determinacy}
We define {\em local determinacy} in a fragment of concurrent games, which slightly generalizes bidding games. We describe the intuition of the definition. Taking a step back, a bidding game has two components: the graph on which the game is played and the budget and tie-breaking mechanism. In a configuration vertex $c = \zug{v, B_1, B_2, s}$ in a bidding game, the triple $\zug{B_1, B_2, s}$ determines the available actions for the two players at $c$. The objective is given by the first element of the configuration vertices; namely, a play gives rise to a path in $V^\omega$ that determines the winner of the game. In {\em R-concurrent} games, we abstract away both elements. Instead of considering the bidding mode of moving, we assume a transducer, denoted $R$, determines the available actions in a configuration vertex. As in bidding games, we allow intermediate vertices between configuration vertices for book-keeping of the state of the transducer. As in bidding games, the objective is determined only by the sequence of configuration vertices that are traversed by a play.

Formally, consider a transducer $R = \zug{A\times A, Q,q_0, \Delta, \Gamma}$, where $\Delta: Q \times A \times A \rightarrow Q$ is a partial function. Let $\lambda: Q\times \set{1,2} \rightarrow 2^A \setminus \set{\emptyset}$ be a function that specifies a set of {\em allowed actions} for each player at every state. For each $a_1 \in \lambda(q, 1)$ and $a_2 \in \lambda(q, 2)$ we require that $\Delta(q, a_1, a_2)$ is defined. Recall that $\Gamma: Q \rightarrow \set{1,2}$. In a transducer that corresponds to a bidding game, each state has the form $\zug{B_1, B_2, s}$, thus it represents the state of the budgets and the state of the tie-breaking mechanism. The allowed actions for \PLi in such a state correspond to the possible bids; namely, for $i \in \set{1,2}$, we have $\lambda(\zug{B_1, B_2, s}, i) = \set{0,\ldots, B_i}$. 

We say that a concurrent game $\G = \zug{A, V, \lambda, \delta, \alpha}$ is {\em $R$-concurrent} for a transducer $R$ if (1) the set of vertices $V$ are partitioned into {\em configuration} vertices $C$ and {\em intermediate} vertices $I$, (2) intermediate vertices do not contribute to the objective, thus for two plays $\pi$ and $\pi'$ that differ only in their intermediate vertices, we have $\pi \in \alpha$ iff $\pi' \in \alpha$, (3) the neighbors of configuration vertices are intermediate vertices and the transition function restricted to configuration vertices is one-to-one, i.e., for every configuration vertex $c$ and two pairs of actions $\zug{a_1,a_2} \neq \zug{a'_1, a'_2}$, we have $\delta(c, a_1, a_2) \neq \delta(c, a'_1, a'_2)$, (4) each intermediate vertex is controlled by one player and its neighbors can either be all intermediate or all configuration vertices, (5) for $v,v' \in V$ with $v \neq v'$ such that $N(v), N(v')  \subseteq I$, we have $N(v) \cap N(v') = \emptyset$, (6) each vertex in $V$ is associated with a state in $R$ with the following restrictions. Suppose $c \in C$ is associated with $q \in Q$. Then, $\lambda(v, i) = \lambda(q, i)$, for $i \in \set{1,2}$. The transducer updates its state after concurrent moves in configuration vertices; namely, for a configuration vertex $c$ and two actions $a_1, a_2 \in A$, let $u=\delta(c, a_1, a_2)$ be an intermediate vertex. Then, the state that is associated with $u$ is $q' = \Delta(q, a_1, a_2)$ and $u$ is controlled by Player~$\Gamma(q')$. The transducer also updates its state between intermediate states; namely, if $u' \in I$ is a neighbor of $u$ and assume \PO controls $u$ and chooses action $a_1$ to proceed from $u$ to $u'$, then $u'$ is associated with $\Delta(q', a_1, a_2)$, for all $a_2 \in A$, and similarly for \PT. Finally, the transducer does not update its state when proceeding from an intermediate vertex to a configuration one; namely, if $c' \in C$ is a neighbor of $u \in I$ and $u$ is associated with $q \in Q$, then $c'$ is associated with $q$.

Bidding games with transducer- and advantage-based tie-breaking are $R$-concurrent. As in the above, for $N \in \Nat$, the states of the transducer $R$ are of the form $\zug{B_1, B_2, s}$, where $B_1 + B_2 = N$ and $s$ is the state of the tie-breaking mechanism. Following a bidding in a configuration vertex, the intermediate vertex is obtained as follows. The budgets are updated by reducing the winning bid from the winner's budget and adding it to the loser's budget, and the state of the tie-breaking mechanism is updated. With transducer-based tie-breaking, we need only one intermediate vertex between two configuration vertices since we use the information from the bidding to update the state of the tie-breaking transducer. In advantage-based tie-breaking, when no tie occurs, a single intermediate vertex is needed since there is no update to the state of the tie-breaking mechanism. In case of a tie, however, a second intermediate vertex is needed in order to allow the player who holds the advantage, the chance to decide whether or not  to use it.

We describe the intuition for local determinacy. Consider a concurrent game $\G$ and a vertex $v$. Recall that it is generally not the case that $\G$ is determined. That is, it is possible that neither \PO nor \PT have a winning strategy from $v$. Suppose \PO has no winning strategy. We say that a transducer admits local determinacy if in every vertex $v$ that is not winning for \PO, there is a \PT action that he can reveal before \PO and stay in a non-losing vertex. Formally, we have the following. 

\begin{defi}[Local determinacy]
\label{def:local-det}
We say that a transducer $R$ admits local determinacy if every concurrent game $\G$ with Borel objective that is $R$-concurrent has the following property. Consider the turn-based game $\G_1$ in which \PO reveals his action first in each position. Since $\alpha$ is Borel, it is a determined game and there is a partition of the vertices to losing and winning vertices for \PO. Then, for every vertex $v \in V$ that is losing for \PO in $\G_1$, there is a \PT action $a_2$ such that, for every \PO action $a_1$, the vertex $\delta(v, a_1, a_2)$ is losing for \PO in $\G_1$.
\end{defi}

We show that locally-determined games are determined by starting with reachability objectives and working our way up to \Muller objectives.
\begin{lem}
\label{lem:safety-det}
If a reachability game $\G$ is $R$-concurrent for a locally-determined transducer $R$, then $\G$ is determined.
\end{lem}
\begin{proof}
Consider a concurrent reachability game $\G = \zug{A, V, \lambda, \delta, \alpha}$ and a vertex $v \in V$ from which \PO does not have a winning strategy. That is, $v$ is losing for \PO in $\G_1$. We describe a winning strategy for \PT from $v$ in $\G$. \PT's strategy maintains the invariant that the set of vertices $S$ that are visited along the play in $\G$, are losing for \PO in $\G_1$. Recall that since we assume intermediate vertices do not contribute to the objective, the target of \PO is a configuration vertex. The invariant implies that \PT wins since there is no intersection between $S$ and \PO's target, and thus the target is never reached. Initially, the invariant holds by the assumption that $v$ is losing for \PO in $\G_1$. Suppose the token is placed on a vertex $u$ in $\G$. Local determinacy implies that \PT can choose an action $a_2$ that guarantees that no matter how \PO chooses, the game reaches a losing vertex for \PO in $\G_1$. Thus, the invariant is maintained, and we are done.
\end{proof}

Next, we show determinacy in parity games by reducing them to reachability games.

\begin{lem}
\label{lem:parity-det}
If a parity game $\P$ is $R$-concurrent for a locally-determined transducer $R$, then $\P$ is determined.
\end{lem}
\begin{proof}
Consider a parity game $\P = \zug{A, V, \delta, \lambda, p}$ that is $R$-concurrent, where $R$ is locally determined. Consider a vertex $v \in V$ from which \PO does not win, and we prove that \PT wins from $v$ in $\P$ (see a depiction of the proof in Figure~\ref{fig:proof}). By Proposition~\ref{prop:det}, for $i \in \set{1,2}$, \PLi wins from $v$ in $\P$ iff he wins from $v$ in $\P_i$ in which he reveals his action first.

\begin{figure}[ht]
\includegraphics[width=\linewidth]{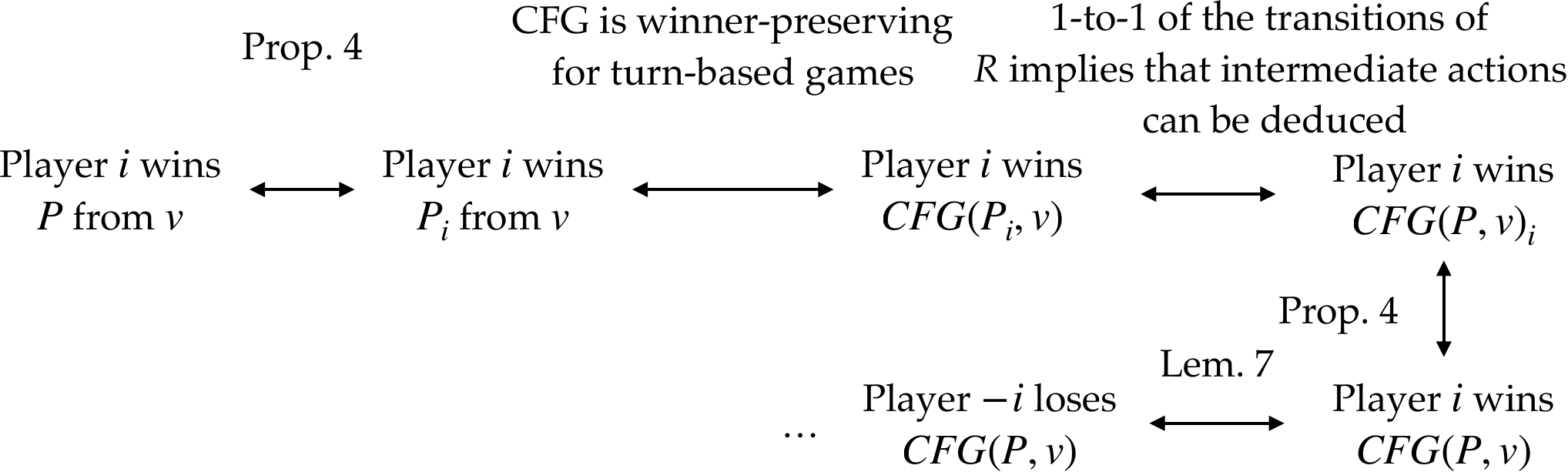}
\caption{A depiction of the proof of Lemma~\ref{lem:parity-det}.}
\label{fig:proof}
\end{figure}

We use a well-known reduction from parity games to reachability games (see for example, \cite{AR17}). The {\em cycle-forming game} that is associated with $\P_i$ and $v$, denoted $CFG(\P_i, v)$, is a reachability game in which we intuitively play from $v$ in $\P_i$ until a cycle is formed. The resulting play is a {\em lasso} $\pi_1 \pi_2$ and \PLi wins iff his objective is met in the infinite play $\pi_1 \pi_2^\omega$. Memoryless determinacy of turn-based parity games \cite{EJ91} implies that \PLi wins from $v$ in $\P_i$ iff he wins from $v$ in $CFG(\P_i, v)$. 

Formally, a vertex in $CFG(\P_i, v)$ records the history of the game in $\P_i$. Recall that in a configuration vertex $c \in V$, \PLi reveals his action first, and, assuming he chooses $a \in A$, the following vertex is $\zug{c,a}$, and its successors are intermediate vertices. Since intermediate vertices are controlled by one of the players and no concurrent moves take place in these vertices, there is no need to add further intermediate vertices. Note that a cycle can only be closed in configuration vertices. Indeed, recall that for $v, v' \in V$, if $N(v), N(v') \subseteq I$, then $N(v) \cap N(v') = \emptyset$. A vertex of $CFG(\P_i, v)$ is a sequence in $(C \times (C \times A) \times I^*)^*$ with no repetitions. Consider a vertex $u = c_1, (c_1, a_1), d^1_1,\ldots, d^1_{n_1},c_2,\ldots,v_k$, where $v_k$ is in $C \cup (C \times A) \cup I$.  If there is an earlier configuration vertex $c_j$ with $v_k = c_j$, then $u$ is a leaf and the winner in it is the winner of the infinite loop as in the above. Otherwise, the player who controls $v_k$ in $\P_i$ controls $u$ and its neighbors are $u \cdot v'$, where $v'$ is a neighbor of $v_k$ in~$\P_i$. 

We apply the same cycle-forming game reduction to the original game $\P$ starting from the vertex $v$. Vertices in $CFG(\P, v)$ are now of the form $(C \times I^*)^*$. Consider a vertex $u = c_1, d^1_1,\ldots, d^1_{n_1}, c_2,\ldots, v_k$. We claim that the resulting game is a reachability game that is $R$-concurrent. Indeed, the vertex $u$ is a configuration vertex in $CFG(\P, v)$ iff $v_k$ is a configuration vertex, and the state in $R$ that $u$ is associated with is the same as $v_k$. If $v_k$ is a configuration vertex, then the allowed actions of the two players in $u$ are the same as in $v_k$. The rest of the construction follows the same lines as the one above. By Lemma~\ref{lem:safety-det}, the game $CFG(\P, v)$ is determined, thus if \PO does not win from $v$ in $CFG(\P, v)$, then \PT wins from $v$ in $CFG(\P, v)$. 

For $i \in \set{1,2}$, we construct $CFG(\P, v)_i$ by requiring \PLi to reveal his choice before \PLni in configuration vertices. Note that $CFG(\P_i, v)$ and $CFG(\P, v)_i$ have a slight technical difference; namely, vertices in $CFG(\P, v)_i$ lack the intermediate vertices in $C \times A$. Since the transition function in $\P$ is one-to-one when restricted to configuration vertices, the vertex between $c_j$ and $d^j_i$ can be uniquely deduced. Thus, \PLi wins from $v$ in $CFG(\P_i, v)$ iff \PLi wins from $v$ in $CFG(\P, v)_i$. 

We combine the reductions: If \PO does not win from $v$ in $\P$, by definition, he loses from $v$ in $\P_1$, thus due to memoryless determinacy in turn-based games, he also loses from $v$ in $CFG(\P_1, v)$ and, due to the equivalence between the games, also in $CFG(\P, v)_1$. Determinacy for reachability games implies that \PT wins from $v$ in $CFG(\P, v)_2$, and going in the other direction, we obtain that \PT wins from $v$ in $\P$, and we are done.
\end{proof}

The proof for \Muller objectives is similar only that we replace the cycle-forming game reduction with a reduction from \Muller games to parity games \cite[Chapter~$2$]{GTW02}. 

\begin{thm}
\label{thm:muller-det}
If a \Muller game $\G$ is $R$-concurrent for a locally-determined transducer $R$, then $\G$ is determined.
\end{thm}

\subsection{The bidding matrix}
Consider a bidding game $\G = \zug{V, E, N, \M,\alpha}$. Recall that $\G$ is $R$-concurrent, where a configuration vertex is of the form $c = \zug{v, B_1, B_2, s}$. The set of allowed actions in $c$ for \PLi is $\set{0, \ldots, B_i}$, for $i \in \set{1,2}$. In particular, there is a natural order on the actions. We think of the possible pairs of actions available in $c$ as a matrix $M_c$, which we call the {\em bidding matrix}. Rows in $M_c$ correspond to \PO bids and columns corresponds to \PT bids. The diagonal that starts in the top-left corner of $M_c$ and follows entries of the form $\zug{j,j}$, for $0 \leq j \leq \min\set{B_1, B_2}$, corresponds to biddings that resolve in a tie. Entries above and below it correspond to biddings that are winning for \PT and \PO, respectively. Consider the turn-based game $\G_1$ in which \PO reveals his bid first. We consider objectives for which turn-based games are determined, thus in $\G_1$, the vertex $\zug{c, b_1, b_2}$ is either winning for \PO or \PT. The entries in $M_c$ are in $\set{1,2}$, where $M_c(b_1, b_2) = 1$ iff the intermediate vertex $\zug{c, b_1, b_2}$ is winning for \PO in $\G_1$. 

For $i \in \set{1,2}$, we call a row or column in $M_c$ an {\em $i$-row} or {\em $i$-column}, respectively, if all its entries are $i$. We rephrase local determinacy in bidding games in terms of the bidding matrix. 

\begin{defi}
\label{def:bidding-local-det}
Consider a bidding game $\G = \zug{V, E, N, \M,\alpha}$. We say that $\G$ is locally determined if for every configuration vertex $c$, the bidding matrix either has a $2$-column or a $1$-row.
\end{defi}

It is not hard to show that Definition~\ref{def:bidding-local-det} implies Definition~\ref{def:local-det}. Consider a bidding game $\G$ in which in each configuration vertex $c$ there is either a $1$-row or a $2$-column in $M_c$. We claim that $\G$ is locally determined. Suppose $c$ is losing for \PO in $\G_1$, we need to show that there is a \PT action (bid) that he can reveal before \PO and that guarantees that the game stays in a losing vertex for \PO. In other words, we need to show that a $2$-column exists. We rule out the possibility of a $1$-row in $M_c$. This is immediate since if there was a $1$-row, \PO could use the corresponding bid, direct the game to a vertex from which he wins, and use the winning strategy from there, contradicting the fact that $c$ is losing for \PO.

\section{Transducer-based tie-breaking}
The determinacy of bidding games with transducer-based tie-breaking depends on the information that is available to the transducer. We start with a negative result.

\begin{thm}
\label{thm:alternate}
Reachability bidding games with alternate tie-breaking are not determined.
\end{thm}
\begin{proof}
Consider the bidding reachability game that is depicted in Fig.~\ref{fig:alternate}. We show that no player has a winning strategy when the game starts from the configuration $\zug{v_0, 1, 1^*}$, thus the token is placed on $v_0$, both budgets equal $1$, and \PT has the tie-breaking advantage. The proof that \PT has no winning strategy is shown in Example~\ref{ex:alternate}. We show that \PO has no winning strategy, thus if he reveals his first bid before \PT, then \PT wins the game. In Fig.~\ref{fig:alternate-proof}, we depict most of the relevant configurations in the game with \PT's strategy in place. Consider the configuration $\zug{v_0, 1, 1^*}$, and we assume \PT reveals his bid after \PO. For example, if \PO bids $0$, \PT bids $0$, wins the bidding since he holds the advantage, and the game proceeds to the configuration $\zug{v_1, 1^*, 1}$. Similarly, if \PO bids $1$, \PT bids $1$, and the game proceeds to $\zug{v_1, 2^*, 0}$. For readability, we omit from the figure some configurations so some configuration have no outgoing edges. It is not hard to show that \PT can force the game from these configurations back to one of the depicted configurations. Thus, when \PO reveals his bids first, \PT can win by forcing the game away from $t$.
\begin{figure}[ht]
\center
\includegraphics[height=3cm]{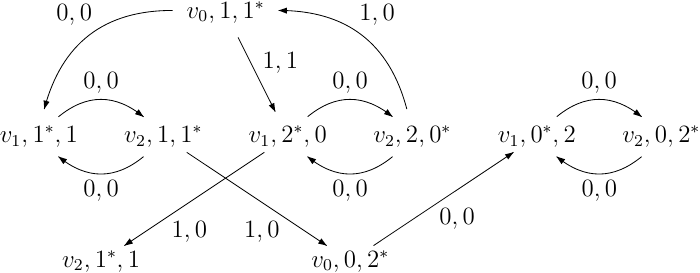}
\caption{Configurations in the game that is depicted in Fig.~\ref{fig:alternate}.}
\label{fig:alternate-proof}
\end{figure}
\end{proof}

We proceed to prove our positive results, namely that bidding games are determined when the information according to which tie-breaking is determined does not include the occurrence of ties. Formally, we define a subclass of tie-breaking transducers.
\begin{defi}
A transducer is {\em un-aware of ties} when its alphabet is $V \times \set{1,2} \times \Nat$, where a letter $\zug{v, i, b} \in V \times \set{1,2} \times \Nat$ means that the token is placed on $v$, \PLi wins the bidding, and his winning bid is $b$. 
\end{defi}

We start with the following lemma that applies to any tie-breaking mechanism. Recall that rows represent \PO bids, columns represent \PT bids, entries on the top-left to bottom-right diagonal represent ties in the bidding, entries above it represent \PT wins, and entries below represent \PO wins.

\begin{lem}
\label{lem:not-on-diag}
Consider a bidding game $\G$ with some tie-breaking mechanism $T$ and consider a configuration $c = \zug{v, B_1, B_2, s}$. Entries in $M_c$ in a column above the diagonal are all equal, thus for bids $b_2 > b_1, b'_1$, the entries $\zug{b_1, b_2}$ and $\zug{b'_1,b_2}$ in $M_c$ are equal. Also, the entries in a row to the left of the diagonal are equal, thus for bids $b_1>b_2, b'_2$, the entries $\zug{b_1, b_2}$ and $\zug{b_1,b'_2}$ in $M_c$ are equal. 
\end{lem}
\begin{proof}
Suppose \PT bids $b_2$. For $b_1, b'_1< b_2$, no matter whether \PO bids $b_1$ or $b'_1$, \PT's budget decreases by $b_2$, thus both the intermediate states $\zug{c, b_1, b_2}$ and $\zug{c, b'_1, b_2}$ are owned by \PT and have the same neighbors. It follows that $\zug{c, b_1, b_2}$ is winning for \PT iff $\zug{c, b'_1, b_2}$ is winning for \PT. The other part of the lemma is dual.
\end{proof}

The next lemma relates an entry on the diagonal with its neighbors. 

\begin{lem}
\label{lem:trans-diag}
Consider a bidding game $\G$ in which tie-breaking is resolved according to a transducer $T$ that is un-aware of ties. Consider a configuration $c = \zug{v, B_1, B_2, s}$. Let $b \in \Nat$. If $\Gamma(s) =1$, i.e., \PO wins ties in $c$, then the entries $\zug{b,b}$ and $\zug{b, b-1}$ in $M_c$ are equal. Dually, if $\Gamma(s) =2$, then the entries $\zug{b,b}$ and $\zug{b-1, b}$ in $M_c$ are equal.
\end{lem}
\begin{proof}
We prove for $\Gamma(s) = 1$, and the other case is dual. Let $c = \zug{v, B_1, B_2, s}$. Note that the neighbors of the intermediate vertices $\zug{c, b, b}$ and $\zug{c, b, b-1}$ are the same. Indeed, \PO is the winner of the bidding in both case, and so his budget decreases by $b$. Also, the update to the state $s$ in $T$ is the same in both cases since $T$ is un-aware of ties. It follows that $\zug{c, b, b}$ is winning for \PO iff $\zug{c, b, b-1}$ is winning for \PO.
\end{proof}

We continue to prove our positive results. 
\begin{thm}
\label{thm:trans-det}
Consider a tie-breaking transducer $T$ that is un-aware of ties. Then, a M\"uller bidding game that resolves ties using $T$ is determined.
\end{thm}
\begin{proof}
We show that transducers that are not aware of ties admit local determinacy, and the theorem follows from Theorem~\ref{thm:muller-det}. See a depiction of the proof in Figure~\ref{fig:trans-det}.

Consider a bidding game $\zug{V, E, \alpha, N, T}$, where $T$ is un-aware of ties, and consider a configuration vertex $c = \zug{v, B_1, B_2, s}$. We show that $M_c$ either has a $1$-row or a $2$-column. We prove for $\Gamma(s) = 1$ and the proof for $\Gamma(s) = 2$ is similar. Let $B = \min\set{B_1, B_2}$. When $B_2 > B_1$, the matrix $M_c$ is a rectangle. Still the diagonal of interest models biddings that result in ties and it starts from the top left corner of $M_c$. The columns $B+1,\ldots, B_2$ do not intersect this diagonal. By Lemma~\ref{lem:not-on-diag}, the entries in each one of these columns are all equal. We assume all the entries are $1$ as otherwise we find a $2$-column. Similarly, if $B_1 > B_2$, we assume that the entries in the rows $B+1,\ldots, B_1$ below the diagonal are all $2$, otherwise we find a $1$-row.

We restrict attention to the $B \times B$ top-left sub-matrix of $M_c$. Consider the $B$-th row in $M_c$. By Lemma~\ref{lem:not-on-diag}, entries in this row that are below the diagonal are all equal, and, since $\Gamma(s) = 1$, they also equal the entry on the diagonal. If all entries equal $1$, then together with the assumption above that entries to the right of the diagonal are all $1$, we find a $1$-row. Thus, we assume all entries below and on the diagonal in the $B$-th row all equal $2$. Now, consider the $B$-th column. By Lemma~\ref{lem:not-on-diag}, the entries above the diagonal are all equal. If they all equal $2$, together with the entry $\zug{B, B}$ on the diagonal and the entries below it, which we assume are all $2$, we find a $2$-column. Thus, we assume the entries in the $B$-th column above the diagonal are all $1$. Next, consider the $(B-1)$-row. Similarly, the elements on and to the left of the diagonal are all equal, and if they equal $1$, we find a $1$-row, thus we assume they are all $2$. We continue in a similar manner until the entry $\zug{1,1}$. If it is $1$, we find a $1$-column and if it is $2$, we find a $2$-row, and we are done.
\end{proof}

We conclude this section by relating the computational complexity of bidding games with turn-based games. Let  $\text{TB}_\alpha$ be the class of turn-based games with a qualitative objective $\alpha$. Let $\text{BID}_{\alpha,\text{trans}}$ be the class of bidding games with transducer-based tie-breaking and objective $\alpha$. The problem TB-WIN$_\alpha$ gets a game $\G \in \text{TB}_\alpha$ and a vertex $v$ in $\G$, and the goal is to decide whether \PO can win from $v$. Similarly, the problem BID-WIN$_{\alpha, \text{trans}}$ gets as input a game $\G \in \text{BID}_{\alpha,\text{trans}}$ with budgets expressed in unary and a configuration $c$ in $\G$, and the goal is to decide whether \PO can win from $c$. 

\begin{thm}
\label{thm:trans-compl}
For a qualitative objective $\alpha$, the complexity of TB-WIN$_\alpha$ and BID-WIN$_{\alpha, \text{trans}}$ coincide when the budgets are given in unary.
\end{thm}
\begin{proof}
In order to decide whether \PO wins in a configuration $c$ in $\G \in \text{BID}_{\alpha,\text{trans}}$, we construct the turn-based game $\G_1$ in which \PO reveals his bids before \PT and solve $\G_1$. The determinacy of $\G$ implies that if \PO does not win $\G_1$, the \PT wins $\G_2$. The size of $\G_1$ is polynomial in $\G$ since the budgets are given in unary.

The other direction is simple: given a turn-based game $\G$, we set the total budgets to $0$, thus all bids result in ties. The tie-breaking transducer resolves ties by declaring the winner in a vertex $v$ to be \PLi if he controls $v$ in $\G$. Clearly, the winner in $\G'$ coincides with the winner in $\G$.
\end{proof}

\section{Random-Based Tie Breaking}
\label{sec:stochas}
In this section we show that bidding games with random-based tie-breaking are determined. A stochastic concurrent game is $\G = \zug{A, V, \lambda, \delta, \alpha}$ is the same as a concurrent game only that the transition function is stochastic, thus given $v \in V$ and $a^1, a^2 \in A$, the transition function $\delta(v, a^1, a^2)$ is a probability distribution over $V$. Two strategies $\sigma_1$ and $\sigma_2$ give rise to a probability distribution $D(\sigma_1, \sigma_2)$ over infinite plays. 

Traditionally, determinacy in stochastic concurrent games states that each vertex is associated with a {\em value}, which is the probability that \PO wins under optimal play \cite{Mar98}. The value is obtained, however, when the players are allowed to use probabilistic strategies. We show a stronger form of determinacy in bidding games; namely, we show that the value exists even when the players are restricted to use deterministic strategies.

\begin{defi}[Determinacy in stochastic games] Consider a stochastic concurrent game $\G$ and a vertex $v \in V$. Let $P_1$ and $P_2$ denote the set of pure strategies for Players~$1$ and~$2$, respectively. For $i \in \set{1,2}$, the value for \PLi, denoted $val_i(\G, v)$, is intuitively obtained when he reveals his strategy before the other player. We define $val_1(\G, v) = \sup_{\sigma_1 \in P_1} \inf_{\sigma_2 \in P_2} \Pr_{\pi \sim D(\sigma_1, \sigma_2)}[\pi \in \alpha]$ and $val_2(\G, v) = \inf_{\sigma_2 \in P_2} \sup_{\sigma_1 \in P_1} \Pr_{\pi \sim D(\sigma_1, \sigma_2)}[\pi \in \alpha]$. We say that $\G$ is determined in $v$ if $val_1(\G, v) = val_2(\G, v)$ in which case we denote the value by $val(\G, v)$. We say that $\G$ is determined if it is determined in all vertices.\hfill\qed
\end{defi}

The key idea in the proof shows determinacy for reachability games that are played on directed acyclic graphs (DAGs, for short). The following lemma shows that the proof for DAGs implies the general case by following an ``unwinding'' argument similar to the one used in the {\em value iteration} algorithm. 
\begin{lem}
Determinacy of reachability bidding games that are played on DAGs implies determinacy of general reachability bidding games.
\end{lem}
\begin{proof}
Let $\G$ be a reachability bidding game with random-based tie breaking and consider a configuration $c$. We claim that $val_1(\G, c) = val_2(\G, c)$. For $i \in \set{1,2}$, recall that $G_i$ is the turn-based stochastic game in which \PLi reveals his bid first in each turn. Trivially, \PLi's value in $\G_i$ at $c$ is $val(\G_i, c)$. For $n \in \Nat$, let $\G^n(c)$ denote the game that starts from $c$ and in which \PO wins iff he reaches the target within $n$ turns. It follows from \cite{Ev55} that the values of $\G^n_i(c)$ converge to the value of $\G_i$ at $c$, thus $val(\G_i, c) = \lim_{n \to \infty} val\big(\G^n_i,(c)\big)$. 

Note that $\G^n(c)$ is a game that is played on a DAG; indeed, the game ends after at most $n$ turns. The game $\G^n_i(c)$ is the game in which \PLi reveals his bid first in each step. The assumption on determinacy of games played on DAGs implies that $val\big(\G^n_1(c)\big) = val\big(\G^n_2(c)\big)$. It thus follows that $val(\G_1, c) = val(\G_2, c)$ since all the elements in the sequence are equal.
\end{proof}

We continue to show determinacy in bidding games on DAGs.
\begin{lem}
Reachability bidding games with random-based tie-breaking that are played on DAGs are determined.
\end{lem}
\begin{proof}
Consider a reachability game  $\G$ that is played on a DAG with two distinguished vertices $t_1$ and $t_2$, which are sinks. There are no other cycles in $\G$, thus all plays end either in $t_1$ or $t_2$, and, for $i \in \set{1,2}$, \PLi wins iff the game ends in $t_i$. The {\em height} of $\G$ is the length of the longest path from some vertex to either $t_1$ or $t_2$. We prove that $\G$ is determined by induction on its height. For a height of $0$, the claim clearly holds since for every $B_1, B_2 \in \Nat$, the value in $t_1$ is $1$ and the value in $t_2$ is $0$. Suppose the claim holds for games of heights of at most $n-1$ and we prove for games of height $n$. 

Consider a configuration vertex $c=\zug{v, B_1, B_2}$ of height $n$. Let $c'$ be a configuration vertex that, skipping intermediate vertices, is a neighbor of $c$. Then, the height of $c'$ is less than $n$ and by the induction hypothesis, its value is well defined. It follows that the value of the intermediate vertices following $c$ are also well-defined: if the intermediate vertex is controlled by \PO or \PT, the value is respectively the maximum or minimum of its neighbors, and if it is controlled by Nature, the value is the average of its two neighbors. 

We claim that $\G$ is determined in $c$ by showing that one of the players has a (weakly) {\em dominant} bid from $c$, where a bid $b_1$ dominates a bid $b'_1$ if, intuitively, \PO always prefers bidding $b_1$ over $b'_1$. It is convenient to consider a variant of the bidding matrix $M_c$ of $c$, which is a $(B_1 + 1) \times (B_2  + 1)$ matrix with entries in $[0,1]$, where an entry $M_c(b_1, b_2)$ represents the value of the intermediate vertex $\zug{c, b_1,b_2}$. Note that \PO, the reachability player, aims to maximize the value while \PT aims to minimize it. We observe some properties of the entries in $M_c$ (see Fig.~\ref{fig:stochas-obs}). 
\begin{itemize}
\item An entry on the diagonal is the average of two of its neighbors, namely $M_c(b, b) = \frac{1}{2} \big(M_c(b-1, b)+M_c(b, b-1)\big)$.
\item As in Lemma~\ref{lem:not-on-diag}, the entries in a column above the diagonal as well as entries in a row to the left of the diagonal, are all equal.
\item For $b_1 > b'_1 > b_2$, we have $M_c(b_1, b_2) \leq M_c(b'_1, b_2)$, since \PO can use the same strategies from $\zug{c, b'_1, b_2}$ as from $\zug{c, b_1, b_2}$. Similarly, for $b_2 > b'_2 > b_1$, we have $M_c(b_1, b_2) \geq M_c(b_1, b'_2)$.
\end{itemize}

\begin{figure}[t]
\begin{minipage}[t]{0.4\linewidth}
\centering
\includegraphics[height=5cm]{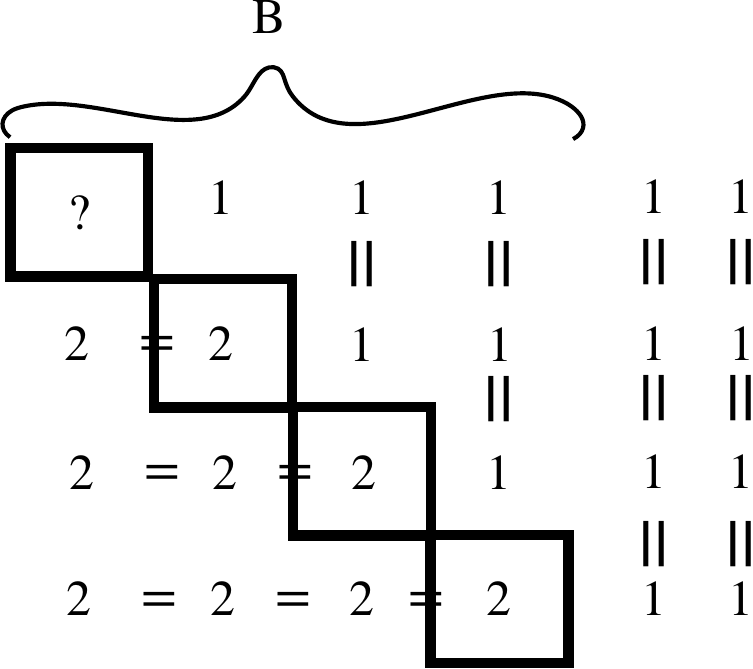}
\caption{A depiction of the contradiction in Theorem~\ref{thm:trans-det} with $B_2 > B_1$.}
\label{fig:trans-det}
\end{minipage}
\hspace{0.05\linewidth}
\begin{minipage}[t]{0.51\linewidth}
\centering
\includegraphics[height=5cm]{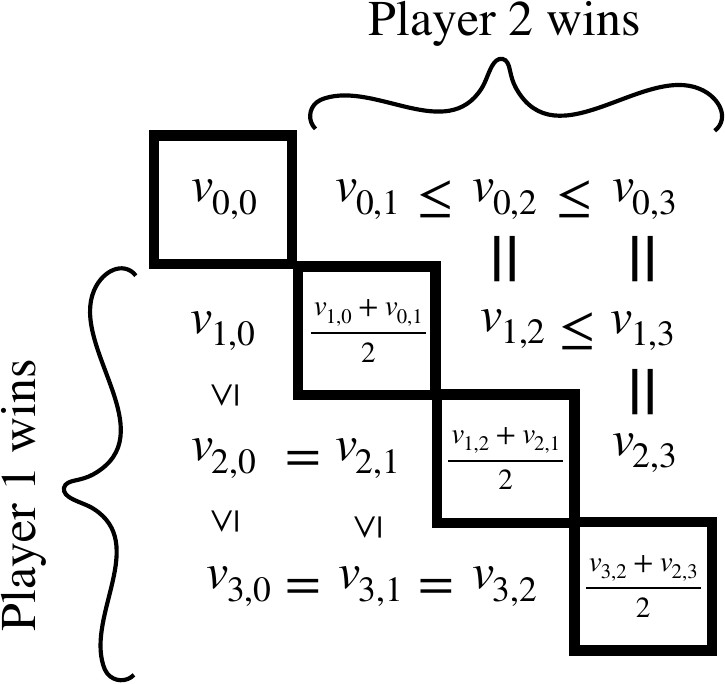}
\caption{Observations on the matrix $M_c$ when resolving ties randomly.}
\label{fig:stochas-obs}
\end{minipage}
\end{figure}

\noindent We show that one of the players has a weakly dominant bid from $c$, where a bid $b_1$ dominates a bid $b'_1$ if for every bid $b_2$ of \PT, we have $M_c(b_1, b_2) \geq M_c(b'_1, b_2)$, and dually for \PT. Consider the bids $0$ and $1$ for the two players. We claim that there is a player for which either $0$ weakly dominates $1$ or vice versa. Assume towards contradiction that this is not the case. Consider the $2\times 2$ top-left sub-matrix of $M_c$ and denote its values $v_{0,0}, v_{0,1}, v_{1,0}$, and $v_{1,1}$. Since $v_{1,1}$ is the average of $v_{0,1}$ and $v_{1,0}$, we either have $v_{0,1} \leq v_{1,1} \leq v_{1,0}$ or $v_{0,1} \geq v_{1,1} \geq v_{1,0}$. Suppose w.l.o.g. that the first holds, thus $v_{0,1} \leq v_{1,0}$. Note that $v_{0,0} < v_{0,1}$, since otherwise the bid $1$ dominates $0$ for \PT. Also, we have $v_{0,0} > v_{1,0}$, since otherwise $0$ dominates $1$ for \PO. Combining, we have that $v_{0,1} > v_{1,0}$, and we reach a contradiction.

Suppose \PO has a dominating row and the case of \PT is dual. To apply the inductive argument, we show two properties: (1) if row $0$ dominates row~$1$, then row $0$ dominates every other row $i$, and (2) if row $1$ dominates row $0$, then column $1$ dominates column $0$ without the first two elements. Property (1) implies that if row $0$ dominates row~$1$, we find a pair of optimal strategies by setting \PO's bid to be $0$ and \PT's bid to be a best response to \PO's bid. Property (2) gives rise to a second inductive argument on the size of $M_c$; namely, if row $1$ dominates row $0$, we can construct a restricted game with the same properties as the original game by removing the first column and row from $M_c$. In the case that row $1$ always dominates row $0$, there are two cases. If the players' budgets are equal, we will end up with a matrix that consists of a unique entry. If \PO's budget is larger than \PT's budget, then we end up with a sub-matrix $M'_c$ that consists of rows that do not intersect the main diagonal of $M_c$, thus the entries in a row $i$ in $M'_c$ are all equal and are larger than those in row $i+1$. Likewise when \PT's budget is larger than \PO's budget. In both cases, one of the players has a weakly dominant strategy.

We conclude the proof by proving the two properties above. We start with Property~(1). Assume row $0$ dominates row $1$. We show that, for $i \geq 1$, row $i$ dominates row $i+1$. Recall that below the diagonal, for every $i \geq 1$ and $j<i$, we have $v_{i,j} \geq v_{i+1,j}$, and above the diagonal, for $j > i$, we have $v_{i, j} = v_{i-1, j}$. We are left with two claims to show; namely, that $v_{i,i} \geq v_{i+1,i}$ and $v_{i,i+1} \geq v_{i+1,i+1}$. Recall that below the diagonal, for $j < i-1$, we have $v_{i,j} = v_{i,j+1}$. Thus, proceeding down from $v_{0,1}$ and then proceeding right, we obtain $v_{0,1} \leq v_{i+1,i}$. Similarly, above the diagonal, by proceeding right from $v_{0,1}$ and then down, we obtain $v_{0,1} \leq v_{i, i+1}$. Since $v_{1,1} = \frac{1}{2} (v_{0,1} + v_{1,0})$ and we assume that $v_{0,1}\geq v_{1,1}$, we have $v_{0,1} \geq v_{1,0}$. Combining the above with $v_{i+1,i+1} = \frac{1}{2} (v_{i,i+1} + v_{i+1,i})$, we obtain $v_{i+1, i} \leq v_{i+1, i+1} \leq v_{i, i+1}$. Observing the previous entry on the diagonal, we note that the same proof shows that $v_{i, i-1} \leq v_{i,i} \leq v_{i-1, i}$. Thus, from $v_{i,i}$, we take one step left, one step down, and one step to the right and obtain $v_{i,i} \geq v_{i,i-1} \geq v_{i+1, i-1} = v_{i+1, i}$, and we are done. 

We proceed to prove Property (2). Assume row $1$ dominates row $0$. As in the above, we have $v_{1,0} \geq v_{1,1}$. Below the diagonal, for every $i \geq 1$, we have $v_{i,0} = v_{i,1}$. 
\end{proof}

Combining the two theorems above, we obtain the following.

\begin{thm}
\label{thm:prob-reach}
Reachability bidding games with random-based tie breaking are determined.
\end{thm}


\section{Advantage-Based Tie-Breaking}
Recall that in advantage-based tie-breaking, one of the players holds the advantage, and when a tie occurs, he can choose whether to win and pass the advantage to the other player, or lose the bidding and keep the advantage. Advantage-based tie-breaking was introduced and studied in \cite{DP10}, where determinacy for reachability games was obtained by showing that each vertex $v$ in the game has a threshold budget $\thresh(v) \in (\Nat \times \set{*})$ such that that \PO wins from $v$ iff his budget is at least $\thresh(v)$, where $n^* \in(\Nat \times \set{*})$ means that \PO wins when he starts with a budget of $n$ as well as the advantage. We show that advantage-based tie-breaking admits local determinacy, thus \Muller bidding games with advantage-based tie-breaking are determined.

Recall that the state of the advantage-based tie-breaking mechanism represents which player has the advantage, thus it is in $\set{1,2}$.
\begin{lemC}[\cite{DP10}]
\label{lem:star-moon}
Consider a reachability bidding game $\G$ with advantage-based tie-breaking.
\begin{itemize}
\item Holding the advantage is advantageous: For $i \in \set{1,2}$, if \PLi wins from a configuration vertex $\zug{v, B_1, B_2, \ni}$, then he also wins from $\zug{v,B_1,B_2,i}$.
\item The advantage can be replaced by a unit of budget: Suppose \PO wins in $\zug{v,B_1,B_2,1}$, then he also wins in $\zug{v,B_1+1,B_2-1,2}$. Suppose \PT wins in $\zug{v,B_1,B_2,2}$, then he also wins in $\zug{v,B_1-1,B_2+1,1}$. 
\end{itemize}
\end{lemC}

\begin{figure}[t]
\begin{minipage}[t]{0.4\linewidth}
\centering
\includegraphics[height=4cm]{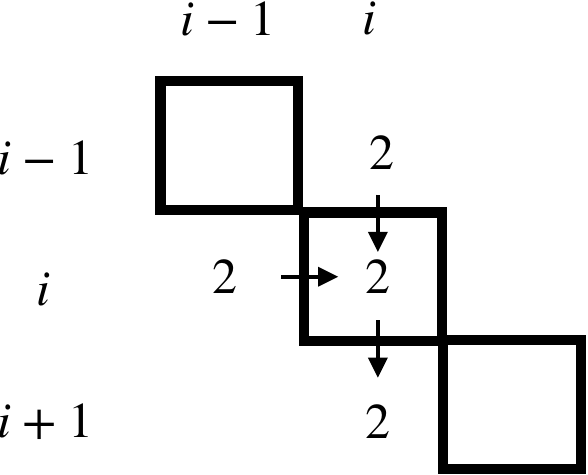}
\caption{A depiction of Lemma~\ref{lem:1}.}
\label{fig:lem-2}
\end{minipage}
\hspace{0.05\linewidth}
\begin{minipage}[t]{0.51\linewidth}
\centering
\includegraphics[height=4cm]{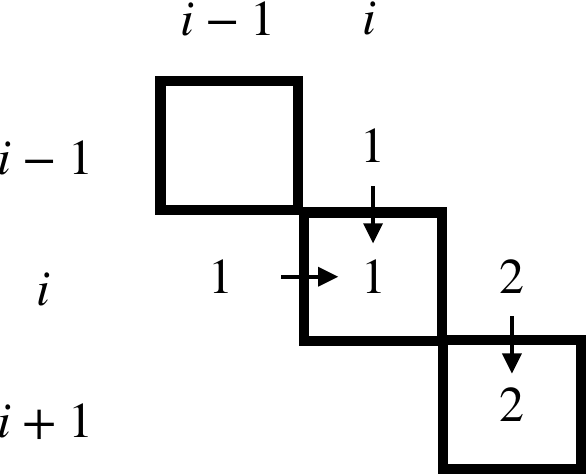}
\caption{A depiction of Lemma~\ref{lem:2}.}
\label{fig:lem-3}
\end{minipage}
\end{figure}

We need two more observations on the bidding matrix, which are depicted in Figs.~\ref{fig:lem-2} and~\ref{fig:lem-3}, and stated in Lemmas~\ref{lem:1} and~\ref{lem:2}.


\begin{lem}
\label{lem:1}
Consider a reachability bidding game $\G$ with advantage-based tie-breaking. Consider a configuration $c = \zug{v, B_1, B_2, 1}$ in $\G$, where \PO has the advantage, and $i \in \set{0,\ldots, B_1}$. Then,
\begin{itemize}
\item If $M_c(i-1, i) = M_c(i, i-1) = 2$, then $M_c(i, i) =2$. 
\item If $M_c(i,i)=2$, then $M_c(i+1, i)=2$. 
\end{itemize}
\end{lem}
\begin{proof}
We start with the first claim. Since both players bid $i$, a tie occurs. Since \PO holds the advantage, there are two cases. In the first case, \PO calls himself the winner and proceeds to a configuration $\zug{v', B_1-i, B_2+i, 2}$. We assume \PT wins from the vertex $\zug{c, i, i-1}$, in which he loses the first bidding. A possible choice of vertex for \PO is $v'$, thus \PT wins from the resulting configuration $\zug{v', B_1-i, B_2+i, 1}$. By Lemma~\ref{lem:star-moon}, \PT also wins $\zug{v', B_1+i, B_2-i, 2}$. In the second case, \PO calls \PT the winner. We assume \PT wins from the vertex $\zug{c, i-1, i}$, in which he wins the first bidding. Let $v'$ be the choice of vertex in a winning strategy, thus the resulting configuration is $\zug{v', B_1-i, B_2+i, 1}$, which is winning for \PT and is the resulting configuration when \PT chooses $v'$ following the tie.

For the second claim, we assume \PT wins in $\zug{c, i, i}$, the vertex that represents a bidding tie. Since \PO has the tie-breaking advantage, \PT wins in particular when \PO calls himself the winner, and the resulting configuration is $\zug{v', B_1 - i, B_2 + i, 2}$. We claim that \PT wins from $\zug{c, i+1, i}$, thus \PO wins the bidding. Let $\zug{v', B_1-(i+1), B_2 + (i+1), 1}$ be the resulting configuration, which by Lemma~\ref{lem:star-moon}, is a \PT winning vertex.
\end{proof}

\begin{lem}
\label{lem:2}
Consider a reachability bidding game $\G$ with advantage-based tie-breaking. Consider a configuration $c = \zug{v, B_1, B_2, 2}$ in $\G$, where \PT has the advantage, and $i \in \set{0,\ldots, B_2}$. Then,
\begin{itemize}
\item If $M_c(i-1,i)=M_c(i, i-1)=1$, then $M_c(i, i)=1$. 
\item If $M_c(i-1, i)=2$, then $M_c(i, i)=2$. 
\end{itemize}
\end{lem}
\begin{proof}
We start with the first claim. Consider a configuration vertex $c = \zug{v, B_1, B_2, 2}$. Since \PO wins when the bids are $i$ and $i-1$, i.e., \PO wins the bidding, there is a configuration vertex $\zug{v', B_1-i, B_2 + i, 2}$ from which \PO wins. Similarly, since \PO wins when the bids are $i-1$ and $i$, he wins no matter which vertex $v''$ \PT chooses to move to, i.e., from configurations of the form $\zug{v'', B_1+i, B_2-i, 2}$. Consider the case that both players bid $i$. \PT has the tie-breaking advantage, thus there are two cases to consider. First, \PT calls himself the winner and chooses the next vertex, thus the proceeding configuration is of the form $\zug{v'', B_1+i, B_2-i, 1}$. Combining the above with Lemma~\ref{lem:star-moon}, \PO wins. Second, \PT calls \PO the winner of the bidding. \PO then chooses $v'$ as in the above, and the following configuration is $\zug{v' B_1-i, B_2+i, 2}$, from which \PO wins.

We continue to the second part of the lemma. Consider the outcome $c'$ in which \PO bids $i-1$ and \PT bids $i$. \PT's budget increases by $i-1$ and he keeps the advantage. On the other hand, consider the outcome $c''$ in which both players bid $i$ and \PT calls \PO the winner. Here, \PT's budget increases by $i$ and the advantage is transferred to \PO. By Lemma~\ref{lem:star-moon}, the advantage can be replaced by a unit of budget. Thus, since \PT wins in $\zug{c', i-1, i}$, he also wins in $\zug{c'', i, i}$. 
\end{proof}

We are ready to prove determinacy.

\begin{thm}
\label{thm:deter-adv}
\Muller bidding games with advantage-based tie-breaking are determined.
\end{thm}
\begin{proof}
Consider a bidding game $\G$ with advantage-based tie-breaking and a configuration $c = \zug{v, B_1, B_2, s}$ in $\G$. We make observations on the entries in $M_c$ above and below the diagonal similar to Thm.~\ref{thm:prob-reach}. Consider the entries above the diagonal. These represent biddings outcomes in which \PT wins. Fixing a \PT bid $b_2$, for any \PO bid $b_1 < b_2$ the outcome of the bidding is the same, i.e., both \PT wins the bidding and the budget update is the same. Thus, entries in a column above the diagonal are all equal. Also, as we proceed right above the diagonal, \PT bids higher and so his updated budget is lower. It follows that if \PO wins in a column $x$, he necessarily wins in every column $x' > x$ to its right. Let $x_2$ denote the first column above the diagonal all of whose entries are $1$. Dually, below the diagonal, entries in rows are equal and as we proceed down, \PO's updated budget is lower. Thus, there is a row, denoted $x_1$, strictly below which all entries are $2$.

\begin{figure}[t]
\centering
\includegraphics[height=4.5cm]{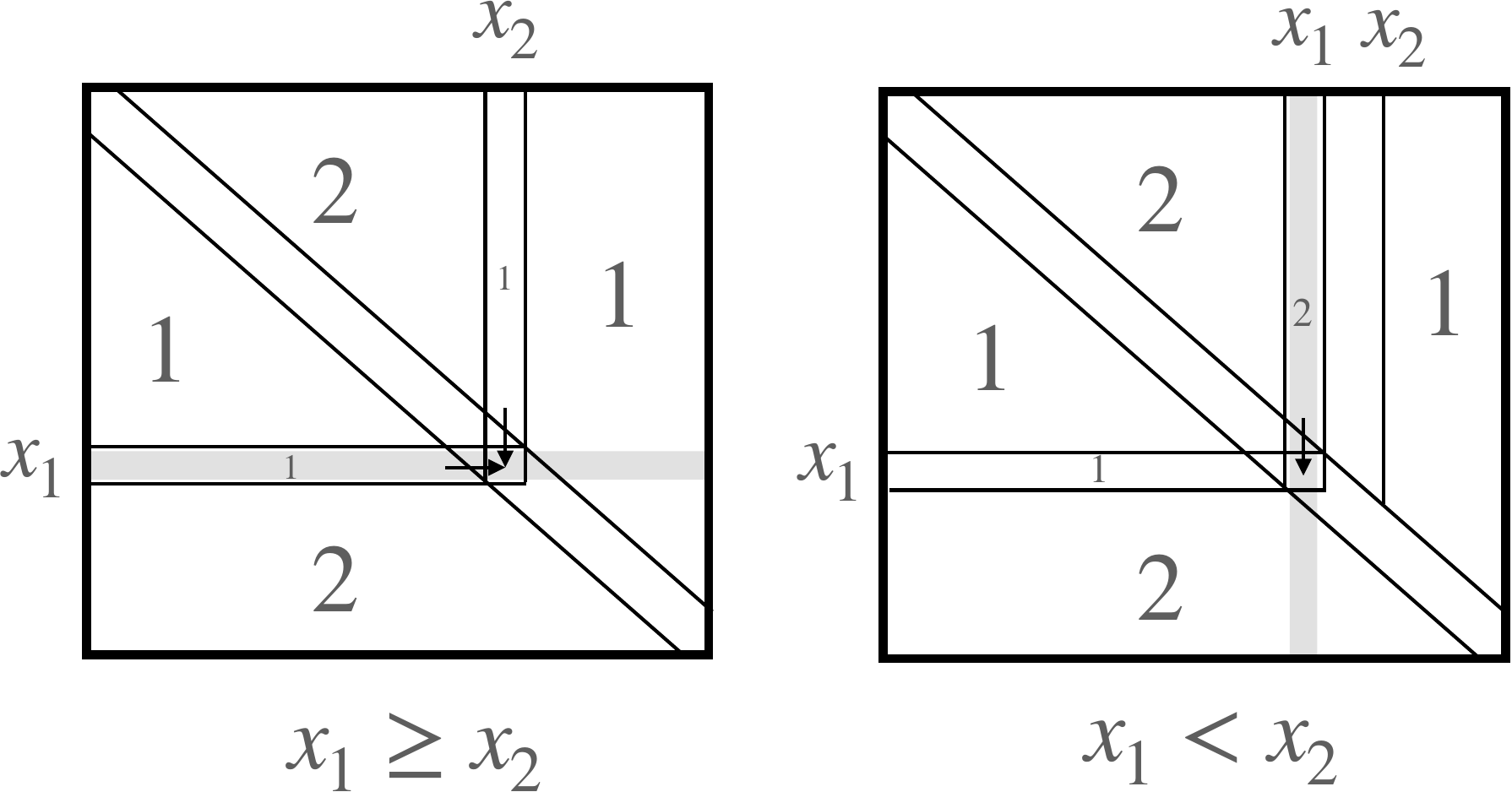}
\caption{A depiction of the cases in which \PT has the advantage in $c$.}
\label{fig:Adv2}
\end{figure}

We distinguish between two cases according to which player has the advantage in $c$. In the first case, \PT has the advantage in $c$ (see a depiction of the proof in Fig.~\ref{fig:Adv2}). We distinguish between two sub-cases. In the first case  $x_2 \leq x_1$. Consider the row $x_1$.  By the definitions of $x_1$ and $x_2$, the entries in the row to the left and to the right of the diagonal are all $1$. In addition, since $x_2 \leq x_1$, the entries in the column $x_1$ above the diagonal are also $1$. Thus, by Lemma~\ref{lem:2}, we have $M_c(x_1, x_1) =1$ and we find a $1$-row. In the second case $x_2 > x_1$. Observe the column $x_1$. By the definitions of $x_1$ and $x_2$ the entries above and below the diagonal are all $2$ and by Lemma~\ref{lem:2}, the entry $x_1$ entry on the diagonal is also $2$, thus we find a $2$-column.

\begin{figure}[t]
\centering
\includegraphics[height=4.5cm]{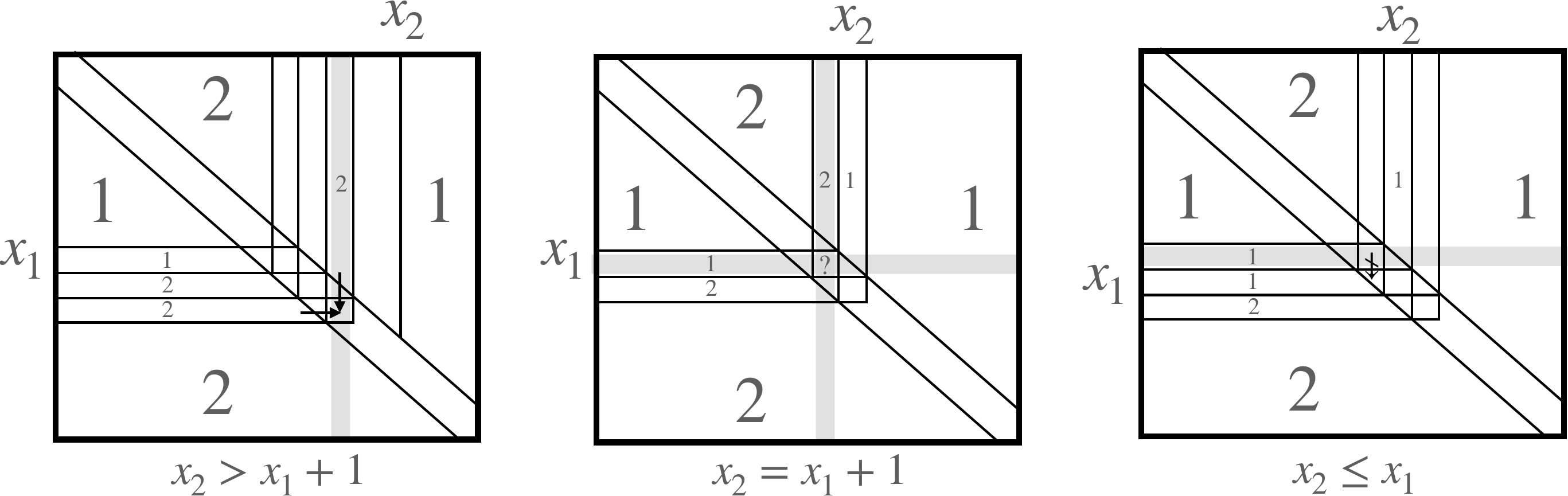}
\caption{A depiction of the cases in which \PO has the advantage in $c$.}
\label{fig:Adv1}
\end{figure}

For the second case, suppose \PO has the advantage (see a depiction of the proof in Fig.~\ref{fig:Adv1}). We distinguish between three sub-cases. In the first case $x_2 > x_1 + 1$. Consider the $(x_1+1)$ column. By the definition of $x_1$ and $x_2$, the entries below and above the diagonal are $2$. Since the entries in the row $(x_1+1)$ to the left of the diagonal are $2$, by Lemma~\ref{lem:1}, the diagonal is also $2$, thus the $(x_1+1)$-column is a $2$-column. In the second case $x_2 = x_1 + 1$. We observe the $x_1$ element of the diagonal. If it is $1$, the $x_1$-row is a $1$-row, and if it is $2$,  the $x_1$-column is a $2$-column. In the third case $x_1 \geq x_2$. Since we have $M_c(x_1, x_1-1)=1$, i.e., the element immediately to the left of the diagonal in the $x_1$ row, the contrapositive of  Lemma~\ref{lem:1} implies that $M_c(x_1-1, x_1-1)=1$. Thus, the $(x_1-1)$-row is a $1$-row, and we are done.
\end{proof}

We turn to study computational complexity of bidding games. Let $\text{BID}_{\alpha,\text{adv}}$ be the class of bidding games with advantage-based tie-breaking and objective $\alpha$, and let BID-WIN$_{\alpha, \text{adv}}$ be the respective decision problem. Recall that TB-WIN$_\alpha$ is the decision problem for turn-based games. 

\begin{thm}
\label{thm:complexity-adv}
For a qualitative objective $\alpha$, the complexity of TB-WIN$_\alpha$ and BID-WIN$_{\alpha, \text{adv}}$ coincide when the budgets are given in unary.
\end{thm}
\begin{proof}
The direction from BID-WIN$_{\alpha, \text{adv}}$ to TB-WIN$_\alpha$ follows from determinacy as in Theorem~\ref{thm:trans-compl}. For the other direction, consider a turn-based game $\G$ and an initial vertex $v_0$. We assume w.l.o.g. that players alternate turns in $\G$. That is, the neighbors of a \PLi vertex $v$ in $\G$ are controlled by \PLni. We construct a bidding game $\G'$ in which the total budgets is $0$. We introduce to $\G$ two new sink vertices $t_1$ and $t_2$, where a play that ends in $t_i$ is winning for \PLi, for $i \in \set{1,2}$. For a \PLi vertex $v$ in $\G$, we add an edge from $v$ to $t_{\ni}$, thus if \PLi has the advantage in $v$, he must use it. Suppose $v_0$ is a \PO vertex in $\G$. It is not hard to show that \PO wins from $v_0$ in $\G'$ when he has the advantage iff he wins from $v_0$ in $\G$.
\end{proof}

\section{Strongly-Connected Games}
Reasoning about strongly-connected games is key to the solution in continuous-bidding infinite-duration games \cite{AHC19,AHI18,AHZ19}. It is shown that in a strongly-connected continuous-bidding game, with every initial positive budget, a player can force the game to visit every vertex infinitely often. It follows that in a strongly-connected \Buchi game $\G$ with at least one accepting state, \PO wins with every positive initial budget. We show a similar result in discrete-bidding games in two cases.

\begin{thm}
\label{thm:SCC}
Consider a strongly-connected bidding game $\G$ in which tie-breaking is either resolved randomly or by a transducer that always prefers \PO. Then, for every pair of initial budgets, \PO can force visiting every vertex in $\G$ infinitely often with probability~$1$. 
\end{thm}
\begin{proof}
Suppose \PO moves whenever a tie occurs and let $v$ be a vertex in the game. \PO follows a strategy in which he always bids $0$ and moves to a vertex that is closer to $v$. For every initial budget of \PT, he wins only a finite number of times. Consider the outcome following the last time \PT wins. Since \PO wins all biddings, in each turn the token moves one step closer to $v$, and thus we visit $v$ every $|V|$ turns, in the worst case. Similarly, when tie-breaking is resolved randomly, the game following the last win of \PT is an ergodic Markov chain in which it is well-known that every vertex is visited infinitely often with probability $1$.
\end{proof}

In \cite{DP10}, it is roughly stated that, with advantage-based tie-breaking, as the budgets tend to infinity, the game ``behaves'' similarly to a continuous-bidding game. The following theorem shows, however, that infinite-duration discrete-bidding games can be quite different from their continuous counterparts. 

\begin{thm}
There is a \Buchi game such that with any pair of initial budgets, \PO wins under continuous-bidding and loses under discrete-bidding.
\end{thm}
\begin{figure}[t]
\centering
\includegraphics[height=1cm]{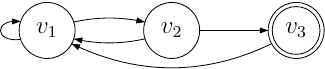}
\caption{A strongly-connected \Buchi game in which \PO wins under continuous bidding and loses under discrete bidding.}
\label{fig:SCC-Buchi}
\end{figure}
\begin{proof}
Consider the game that is depicted in Fig.~\ref{fig:SCC-Buchi} with the initial vertex $v_1$. Since the game is strongly-connected and it has an accepting vertex, by \cite{AHC19}, \PO wins under continuous bidding with any positive initial budget. We proceed to study the game under discrete bidding. Suppose \PO's budget is $B_1 \in \Nat$ and assume wlog that \PT's budget is $0$ (\PT can always ignore excess funds and play as if his initial budget is $0$). \PT always bids $0$, uses the advantage when he has it, and, upon winning, stays in $v_1$ and moves from $v_2$ to $v_1$. Note that in order to visit $v_3$, \PO needs to win two biddings in a row; in $v_1$ and $v_2$. Thus, in order to visit $v_3$, he must ``invest'' a unit of budget, meaning that the number of visits to $v_3$ is bounded by $B_1$, and in particular \PO cannot force infinite many visits to $v_3$, thus he loses the game.
\end{proof}

\section{Discussion and Future Work}
We study discrete-bidding infinite-duration bidding games and identify large fragments of bidding games that are determined. Bidding games are a subclass of concurrent games. We are not aware of other subclasses of concurrent games that admit determinacy. We find it an interesting future direction to extend the determinacy we show here beyond bidding games. Weaker versions of determinacy in fragments of concurrent games have been previously studied \cite{Rou18}.

We focused on bidding games with ``Richman'' bidding, i.e., the winner of the bidding pays the other player, and it is interesting to study other bidding games with other bidding rules. Bidding reachability games with {\em all-pay} bidding in which both players pay their bid to the bank were studied with continuous bidding \cite{AIT20} as well as with discrete Richman-all-pay bidding \cite{MWX15} in which both players pay their bid to the other player. In addition, it is interesting to study discrete-bidding games with quantitative objectives and non-zero-sum games, which were previously studied only for continuous bidding \cite{AHC19,AHI18,MKT18}.

This work belongs to a line of works that transfer concepts and ideas between the areas of formal verification and algorithmic game theory \cite{NRTV07}. 
Examples of works in the intersection of the two fields include logics for specifying multi-agent systems \cite{AHK02,CHP10,MMPV14}, studies of equilibria in games related to synthesis and repair problems \cite{CHJ06,Cha06,FKL10,AAK15}, non-zero-sum games in formal verification \cite{CMJ04,BBPG12}, and applying concepts from formal methods to {\em resource allocation games} such as rich specifications \cite{AKT16}, efficient reasoning about very large games \cite{AGK17,KT17}, and a dynamic selection of resources \cite{AHK16}.

\bibliographystyle{plain}
\bibliography{../ga}
\end{document}